\documentclass[%
 reprint,
 superscriptaddress,
 amsmath,amssymb,
 aps,
]{revtex4-2}

\usepackage{graphicx}% Include figure files
\usepackage{float}
\usepackage{dcolumn}% Align table columns on decimal point
\usepackage{bm}% bold math
\begin{document}

%\preprint{APS/123-QED}

%\title{Approaching exact ground states of the many-electron Schr\"odinger equation with neural wavefunctions}% Force line breaks with \\
\title{Scaling universal Fermi network toward ground states: \\ A diffusion-Monte-Carlo assessment}

\author{Yu-Sheng Li}
\thanks{These authors contributed equally to this work.}
\affiliation{Institute of Physics, Academia Sinica, Taipei 11529, Taiwan}

\author{Saskia Poldmaa}
\thanks{These authors contributed equally to this work.}
\affiliation{Department of Physics, Massachusetts Institute of Technology, Cambridge, Massachusetts 02139, USA}

\author{Tzen Ong}
\affiliation{Institute of Physics, Academia Sinica, Taipei 11529, Taiwan}

\author{Ahmed Abouelkomsan}
\affiliation{Department of Physics, Massachusetts Institute of Technology, Cambridge, Massachusetts 02139, USA}

\author{Tay-Rong Chang}
\affiliation{Department of Physics, National Cheng Kung University, Tainan 70101, Taiwan}

\author{Hsin Lin}%
\affiliation{Institute of Physics, Academia Sinica, Taipei 11529, Taiwan}

\author{Liang Fu}
\email{liangfu@mit.edu}
\affiliation{Department of Physics, Massachusetts Institute of Technology, Cambridge, Massachusetts 02139, USA}



\begin{abstract}
In this work, we show that Fermi Sets---a provably universal neural network architecture for fermionic wavefunctions---can be systematically scaled up to find interacting ground states  through energy minimization in a variational Monte Carlo framework. By further performing fixed-phase diffusion Monte Carlo (DMC) on the optimized neural network wavefunction, we demonstrate that as the network size increases, the variational energy systematically decreases while the energy improvement from DMC collapses monotonically to zero, indicating convergence to the ground state. We illustrate the scaling of Fermi Sets accompanied by the DMC assessment for interacting electrons in jellium and in a quantum dot under high magnetic fields.
%The quantum dot problem illustrates that NN-VMC performs well on complex-valued wavefunctions. Using the 2D UEG, we validate the robustness of this framework across larger system sizes subject to periodic boundary conditions.

\end{abstract}

\maketitle
\begin{figure*}
\includegraphics[width=\textwidth]{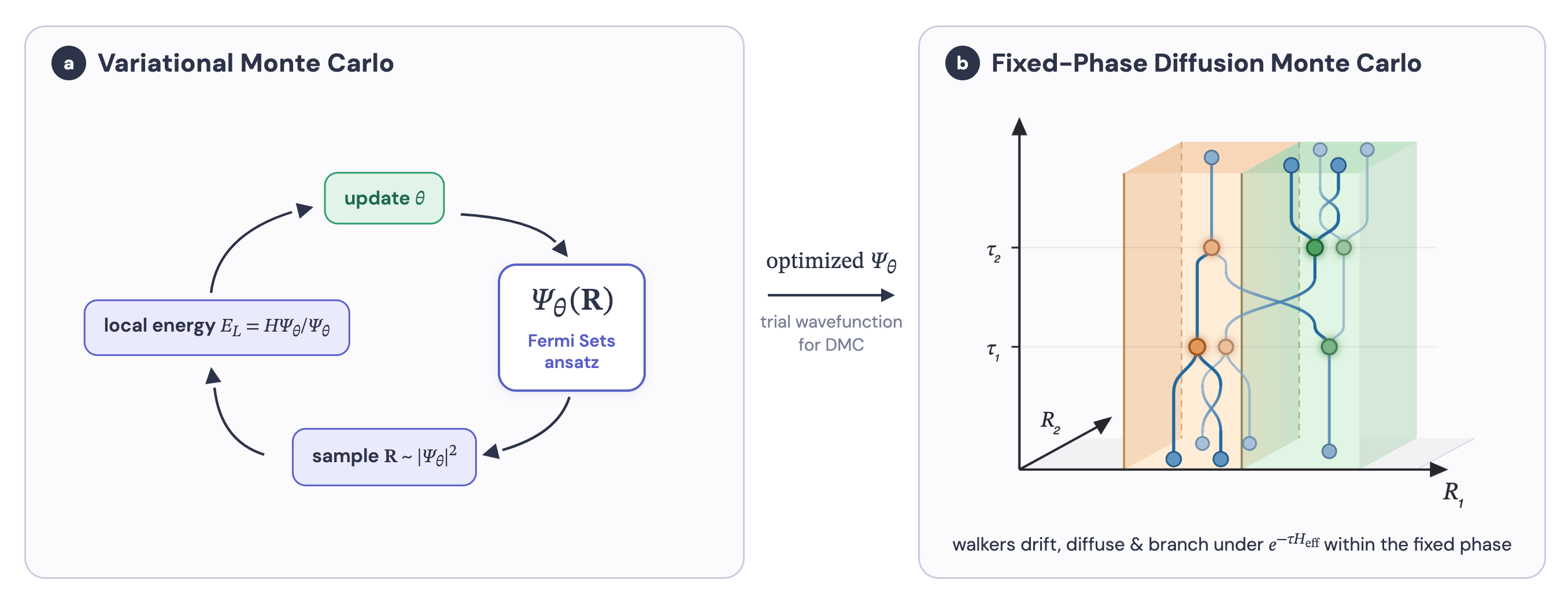}
\caption{\label{fig:wide} Two-stage framework for optimizing and assessing Fermi Sets wavefunctions. (a) Variational optimization of the Fermi Sets ansatz [Eq.~(\ref{eq:fermisets})] by energy minimization (Sec.~\ref{sec:fermisets}). (b)~Fixed-phase diffusion Monte Carlo (Sec.~\ref{sec:FPDMC}): walkers drift and diffuse through configuration space (axes $R_1$, $R_2$) at the learned phase $\Phi_\theta(\mathbf{R})$, depleting in high-energy regions (orange) and proliferating in low-energy ones (green). The resulting gap $\Delta E=E_{VMC}-E_{DMC}\geq 0$ measures the residual amplitude error.
}
\end{figure*}

\section{\label{sec:intro} Introduction}
Solving the electronic Schr\"odinger equation---which governs the behavior of atoms, molecules and materials---is a central goal and challenge in quantum condensed matter physics, quantum chemistry and materials science. The fundamental difficulty of this problem is twofold. First, the Hilbert space of continuum systems is infinite-dimensional even for a finite number of electrons, $N$, i.e., the functional space of all possible wavefunctions $\psi(\bm r_1,...,\bm r_N)$ is unbounded. Second, Fermi statistics requires that the electron wavefunction must be antisymmetric under particle exchange. The functional space of antisymmetric functions of $N$ particle coordinates is a highly complex manifold. 

Traditional methods begin with a finite set of one-electron basis functions, on which a many-body expansion is built. Full configuration interaction (FCI) solves the electronic Schr\"odinger equation exactly within a chosen basis. However, the number of many-body configurations grows factorially with the number of electrons and orbitals. Even for modest molecules in a standard basis set, the dimension of the Hamiltonian matrix can quickly scale to billions or even trillions of configurations, and the accuracy of FCI results remains limited by the finite size of the one-electron basis set. Configuration interaction and coupled cluster truncate the expansion at fixed excitation level; 
their accuracy is limited by both the incomplete single-particle basis and the many-body truncation. 
Diffusion Monte Carlo operates directly in continuous real space~\cite{Ceperley1980,Foulkes2001}, but requires a trial wavefunction whose nodes are fixed by design, yielding an uncontrolled fixed-node error~\cite{Anderson1975,Reynolds1982}. 
%While each of these methods finds applications in certain domain, for a long time there was no general method for solving many-electron Schr\"odinger equation in continuous space that is  numerically efficient, and systematically improvable with compute.  
%None of these established methods delivers the combination of a basis-free representation and exact nodal structure.    

Recently, neural networks, especially those with self attention~\cite{vonGlehn2023Psiformer, Geier2025Attention}, have emerged as a powerful  variational method for solving many-electron Schr\"odinger equation in continuous space~\cite{Pfau2020FermiNet, Cassella2023HEG, Teng2025FQH, khach26}, without using basis sets or discretization. Instead, a  neural network directly generates antisymmetric many-body wavefunctions, which are then optimized to minimize the variational energy with respect to the network parameters. %Recent works have shown that neural network wavefunctions can accurately find strongly correlated Fermi liquids and Wigner crystals~\cite{Cassella2023HEG}, and self-attention neural networks~\cite{Geier2025Attention} can even capture highly entangled and topologically ordered states~\cite{Teng2025FQH, https://arxiv.org/abs/2507.13322}.

Compared to handcrafted trial wavefunctions, neural wavefunctions have the potential advantage of being more expressive---and more importantly---systematically improvable upon increasing the size of the neural network. Indeed, the universal approximation theorem---a key foundation of machine learning---states that a feedforward neural network at sufficient size  can approximate any continuous function to arbitrary accuracy. For the purpose of solving many-electron ground states, it is important to develop Fermi networks that exactly enforce the antisymmetry by design and can approximate any continuous antisymmetric function \cite{Chen2025, Fu2026FermiSets}. 

Very recently, Ref.~\cite{Fu2026FermiSets} has demonstrated that a {\it complex-valued}  neural network architecture  built from a small number of Slater determinants weighted by expressive symmetric functions---``Fermi Sets''---is a {\it universal} approximator of continuous antisymmetric functions. By design, Fermi Sets can capture arbitrary nodal structure, which is crucial for achieving universal representational power.   
Therefore, Fermi Sets wavefunctions can in principle achieve arbitrary accuracy as the network capacity increases, pushing the variational energy to the exact limit. 
In this sense, Fermi Sets architecture provides a universal and asymptotically-exact solver of the electronic Schr\"odinger equation.    

While Fermi Sets has achieved accurate results on several many-electron problems~\cite{Fu2026FermiSets, Tim26}, 
the feasibility of scaling Fermi Sets systematically toward exact ground states has not been explored. An immediate challenge is how to assess the deviation of an optimized neural wavefunction from the exact ground state, which is in general unknown. This issue becomes particularly pressing when the variational energy of the neural wavefunction beats all existing benchmarks, such as full configuration interaction within a finite basis set and traditional quantum Monte Carlo~\cite{Pfau2020FermiNet,Teng2025FQH,Fu2026FermiSets}. As the community increasingly deploys neural networks to solve ground state of many-electron systems, there is an urgent and critical need for rigorous assessments of the accuracy of neural wavefunctions.

In this work, we introduce a general computational framework that combines complex-valued neural-network wavefunctions with fixed-phase diffusion Monte Carlo (FPDMC)~\cite{Ortiz1993} to both improve and assess them. Whereas previous combinations of neural wavefunctions with projector Monte Carlo have relied on the real-valued fixed-node approximation~\cite{Wilson2021DMC,Ren2023DMC}, the fixed-phase constraint applies to arbitrary complex-valued trial wavefunctions, reduces to the fixed-node constraint in the real limit, and remains valid when time-reversal symmetry is broken~\cite{Ortiz1993,Bolton1996}. By using the optimized neural wavefunction as the trial state of an imaginary-time FPDMC projection (Figure~\ref{fig:wide}), we eliminate residual amplitude errors while holding the learned phase structure fixed. The resulting reduction of the energy, $\Delta E = E_{\text{VMC}} - E_{\text{DMC}} \geq 0$, provides a rigorous, quantitative metric for the neural wavefunction's quality.

% Fermi Sets is one member of a growing family of expressive neural-network ansatzes; another is the Psiformer~\cite{vonGlehn2023Psiformer}, and neural wavefunctions equipped with periodic input features have been applied successfully to the electron gas~\cite{Cassella2023HEG, Wilson2023HEG, Smith2024UEG}. We deliberately employ two structurally distinct members of this family on two physically distinct systems---Fermi  Sets for a quantum dot in a magnetic field, a Psiformer adapted to periodic boundary conditions for the uniform electron gas. This pairing carries no claim that either ansatz is uniquely suited to its system; its purpose is to demonstrate that the framework is universal, tied to neither a particular architecture nor a particular physical setting.

We first apply the framework to a parabolic quantum dot in a perpendicular magnetic field, where the exact ground state is intrinsically complex-valued, lies beyond the reach of fixed-node projection, and  for a few electrons, can be checked against multi-Landau-level exact diagonalization. Using Fermi Sets variational Monte Carlo, we find that for the quantum dot problem, the residual gap $\Delta E$ decreases monotonically %from $10^{-4}$  
to as small as $\sim 10^{-6}$ of the ground state energy as the network capacity grows, while the variational energy agrees with the exact-diagonalization benchmark. As a second example, we apply Fermi Sets to the strongly correlated two-dimensional uniform electron gas ($N=16$ electrons at $r_s=30$), and %show that 
find that the residual gap $\Delta E$ collapses to $\sim 2\times10^{-4}$ of the ground state energy. %, by nearly a factor of three, as the network is scaled from $2.5\times10^{4}$ to $5\times10^{4}$ parameters. 
In both cases, the collapse of $\Delta E$ indicates convergence of the optimized Fermi Sets wavefunction to the true ground state.

Remarkably, using the same determinant-based Fermi Sets architecture, we obtain nearly exact ground states in diverse settings. These include  integer quantum Hall and composite fermion states in a quantum dot under magnetic fields, as well as a strongly correlated Fermi liquid at zero magnetic field. Our findings support the notion that Fermi Sets is a universal and asymptotically exact ground state solver for many-electron systems.

\section{\label{sec:methods}Methods}

Our computational framework consists of two distinct stages: the variational optimization of a complex-valued neural wavefunction, followed by an imaginary-time projection subject to the fixed-phase approximation. %Two ansatzes are employed, matched to the two physical settings of Sec.~\ref{sec:results}: 
For variational optimization, both problems employ the Fermi Sets wavefunction (Sec.~\ref{sec:fermisets}). This is followed by a DMC projection (Sec.~\ref{sec:FPDMC}). Wavefunctions are evaluated in the log domain for numerical stability, from which the amplitude $|\Psi_{\theta}|$, the continuous phase $\Phi_{\theta}(\mathbf{R}) = \arg\Psi_{\theta}(\mathbf{R})$, and their gradients follow analytically by automatic differentiation.

\subsection{\label{sec:fermisets} Fermi Sets wavefunction for the quantum dot}
\begin{figure}[t]
\includegraphics[width=\columnwidth]{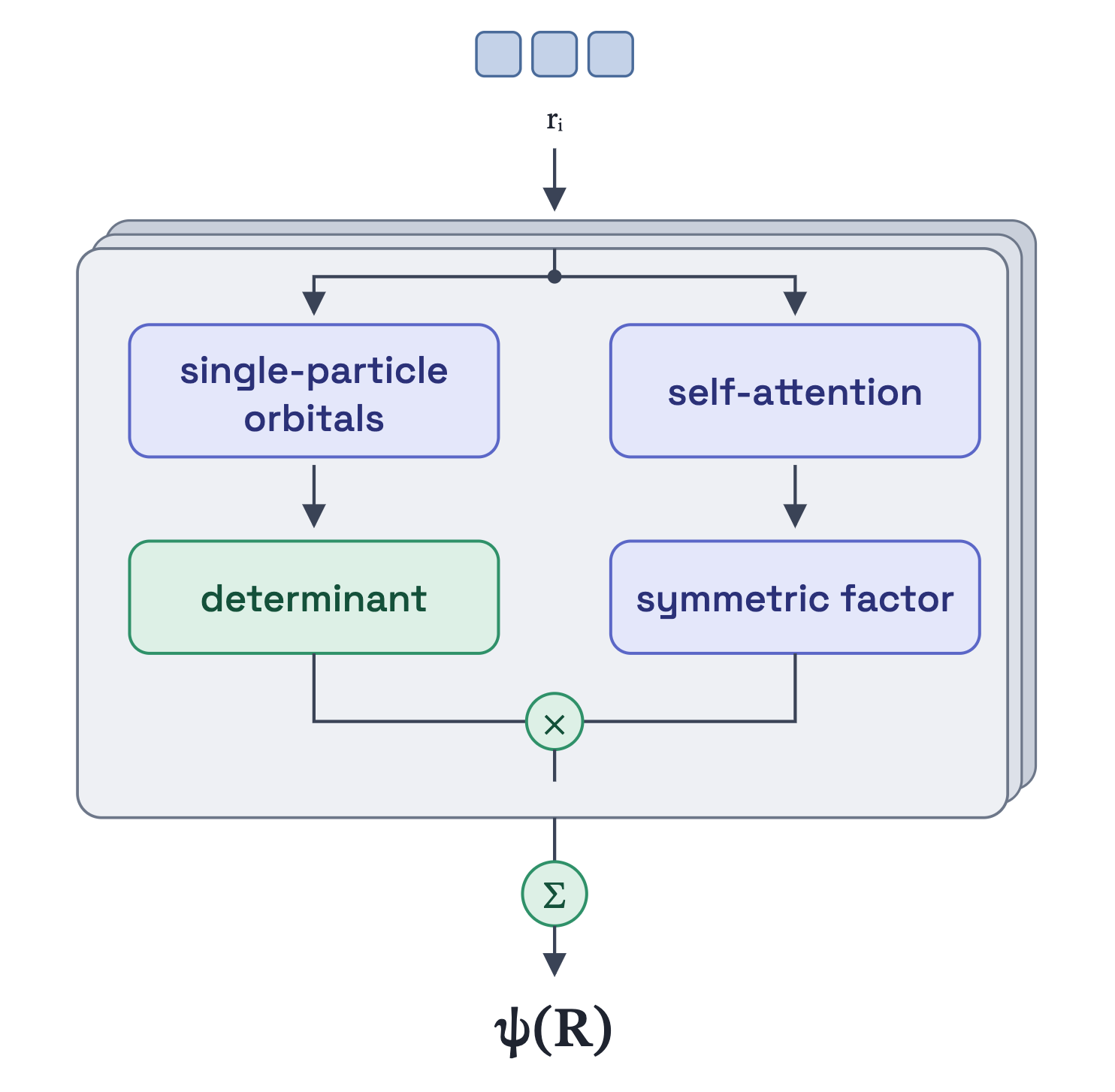}
\caption{\label{fig:fermisets}Architecture of the Fermi Sets wavefunction $\Psi_{\theta}(\mathbf R)$ (Sec.~\ref{sec:fermisets}). The wavefunction is a sum over $N_d$ products of determinants supplying fermionic antisymmetry, and symmetric factors carrying the electron-electron correlations.}

\end{figure}
Fermi Sets~\cite{Fu2026FermiSets} is a complex-valued neural-network wavefunction that represents the many-electron state as a linear combination of a small number $N_d$ of Slater determinants with learnable orbitals, weighted by symmetric functions. Concretely, the Fermi Sets wavefunction, augmented with an optional envelope and Jastrow factor, reads as follows,
\begin{equation}
\Psi_{\theta}(\mathbf{R}) \;=\; e^{J_\theta(\mathbf{R})} e^{\mathcal E(\mathbf{R})}
\sum_{k=1}^{N_d} \Omega_k(\mathbf{R})\,
\det\!\big[\phi_i^{\,k}(\mathbf{r}_j)\big]_{i,j=1}^{N},
\label{eq:fermisets}
\end{equation}
where $\mathbf R=(\mathbf r_1, ... \mathbf r_N)$. 

When the orbitals $\phi_i^{\,k}(\mathbf{r})$ and the weights $\Omega_k(\mathbf{R})$ are allowed to be fully general complex-valued functions, Fermi Sets can provably approximate any continuous antisymmetric function in two dimensions using only $N_d=2$ determinants~\cite{Fu2026FermiSets}. In practice, the single-particle orbital $\phi_i^{\,k}$ is implemented with a multilayer perceptron (MLP), while the many-body symmetric function $\Omega_k(\mathbf{R})$ is implemented with a permutation invariant neural network, such as 
Deep Sets~\cite{Zaheer2017DeepSets} or Transformer~\cite{Vaswani2017Attention} with sum pooling, which are universal approximators of symmetric functions. Figure~\ref{fig:fermisets} illustratses the Fermi Sets construction. Details about our implementation can be found in Appendix~\ref{app:backbone}.

Here, we also include an optional Jastrow factor $e^{J_\theta(\mathbf R)}$ — while it can in
principle be absorbed into the symmetric weights $\Omega_k(\mathbf R)$, we keep a Jastrow factor for the quantum dot, as it aids with convergence. As is customary for Coulomb problems, we use the Pad\'e Jastrow~\cite{Ceperley}
\begin{equation}
J_\theta(\mathbf R)=\sum_{i<j}\frac{\alpha\,r_{ij}}{1+\beta\,r_{ij}},
\label{eq:jastrow}
\end{equation}
where $r_{ij}$ is the absolute distance between electrons i and j. The cusp coefficient $\alpha$ is fixed at $\alpha=\tfrac13$ for spin-polarized electrons in two
dimensions, while the range parameter $\beta>0$ is learnable. This form of the Jastrow captures the correct behavior in both limits: as $r_{ij}\to0$ it reduces to
$\mathcal J_\theta\to \tfrac13 r_{ij}$, reproducing the exact linear cusp of a Coulomb problem~\cite{Ceperley}, and as $r_{ij}\to \infty$, it stabilizes to a constant $\tfrac \alpha \beta$.

Lastly, for the quantum dot we include an envelope in the ansatz, which biases the wavefunction to decay to zero at large radius. The envelope guarantees normalizability, and spares the network from having to learn the asymptotic tail.  
We use the Gaussian envelope $e^{\mathcal E}$,
\begin{equation}
\mathcal E(\mathbf R)=-\sigma\sum_{j=1}^{N}|z_j|^2 ,\qquad z_j=x_j+iy_j,
\label{eq:env}
\end{equation}
where $\sigma$ is a learnable parameter. %Like the Jastrow, the envelope is a convenience, and could be easily be replaced by  Fermi Set~\cite{Fu2026FermiSets}.

Because both the orbitals and the symmetric factors are natively complex-valued, the Fermi Sets wavefunction allows for intrinsically complex-valued wavefunctions that describe electrons in a magnetic field. %For the quantum dot, network capacity is scaled by varying the width of the hidden layers, allowing for a range of parameter counts $N_p \in [8.5\times10^{4},\, 2.0\times10^{5}]$, while keeping the number of determinants at the theoretical minimum of $N_d=2$.

For the electron gas, coordinates enter the network only through periodic features: positions and electron--electron differences are mapped to fractional coordinates $\mathbf{s} = \Lambda^{-1}\mathbf{r}$, with $\Lambda = (\mathbf{L}_1, \mathbf{L}_2)$ the lattice matrix of the simulation cell (Appendix~\ref{app:ueg}), and embedded as $\bigl(\sin 2\pi \mathbf{s}, \cos 2\pi \mathbf{s}\bigr)$, while scalar distances are replaced by the periodic norm of Ref.~\cite{Cassella2023HEG}, which reduces to the Euclidean distance at short range. The ansatz is thus exactly periodic under translation of any electron by a lattice vector. The envelope and Jastrow factor are not used.
%: normalizability is automatic on the torus, and   short-range correlations, including the Coulomb cusp, are left to the network.
 
\subsection{\label{sec:FPDMC} Fixed-Phase Diffusion Monte Carlo}
Diffusion Monte Carlo isolates the ground state by evolving the imaginary-time Schr\"odinger equation: the propagator $e^{-\tau(\hat H - E_0)}$ filters out all excited-state components of the trial state as $\tau \to \infty$~\cite{Foulkes2001}. For fermions, an unconstrained projection suffers from the exponential signal-to-noise decay of the sign problem; the standard remedy for real-valued trial states is the fixed-node approximation~\cite{Anderson1975,Reynolds1982}, which requires a real trial state. However, for complex wavefunctions we employ the fixed-phase method of Ortiz, Ceperley, and Martin~\cite{Ortiz1993}, which applies to the arbitrary complex-valued ansatzes used here and contains the fixed-node constraint as the special case of a real trial state. The trial wavefunction is separated into its amplitude and phase,
\begin{equation}
\Psi_{\theta}(\mathbf{R}) = |\Psi_{\theta}(\mathbf{R})|\, e^{i \Phi_{\theta}(\mathbf{R})},
\label{eq:polar}
\end{equation}
and the phase of the projected state is constrained to that of the trial state for all imaginary time,
\begin{equation}
\Phi(\mathbf{R}, \tau) = \Phi_{\theta}(\mathbf{R}).
\label{eq:phase}
\end{equation}
Under this constraint the real, nonnegative amplitude evolves in imaginary time under the effective Hamiltonian~\cite{Ortiz1993} in Hartree atomic units reads
\begin{equation}
\hat{H}_{\text{eff}} = -\frac{1}{2}\sum_{i=1}^{N} \nabla_i^{2} + V(\mathbf{R}) + \frac{1}{2} \sum_{i=1}^N \big| \nabla_i \Phi_{\theta}(\mathbf{R}) - \mathbf{A}(\mathbf{r}_i) \big|^2,
\label{eq:heff}
\end{equation}
where $\mathbf{A}$ is the vector potential of the applied magnetic field. The phase gradient plays the role of an effective vector potential and, combined with $\mathbf{A}$, acts as a repulsive potential that expels walkers from regions of rapidly winding phase. The random walk is therefore free of the sign problem, while the exact antisymmetry of the state is preserved: it is carried by the fixed complex phase, which shifts by $\pi$ under any transposition of two electrons while $|\Psi_{\theta}|$ is exchange symmetric.
 
The critical advantage of FPDMC in this context is its strict variational structure. The asymptotic energy of the projection, the fixed-phase energy $E_{\text{FP}}$, is by definition the minimum of the Rayleigh quotient of $\hat H$ over all states of the form $\rho(\mathbf{R})\, e^{i\Phi_{\theta}(\mathbf{R})}$ with nonnegative, exchange-symmetric amplitude $\rho$, i.e., it is the lowest energy physically permissible for the specific phase manifold $\Phi_{\theta}$ learned by the network. Two inequalities follow immediately:
\begin{equation}
E_0 \;\le\; E_{\text{FP}} \;\le\; E_{\text{VMC}}.
\label{eq:chain}
\end{equation}
The right inequality holds because $\rho = |\Psi_{\theta}|$ is itself an admissible amplitude; the left inequality holds because every state $\rho\, e^{i\Phi_{\theta}}$ is a valid antisymmetric wavefunction, so the Rayleigh--Ritz principle applies, with equality if and only if $\Phi_{\theta}$ coincides with the phase of an exact ground state~\cite{Ortiz1993}. Consequently, the total variational error decomposes exactly into two nonnegative parts,
\begin{equation}
E_{\text{VMC}} - E_0 =
\underbrace{\big(E_{\text{VMC}} - E_{\text{FP}}\big)}_{\Delta E \,\ge\, 0}
+ \underbrace{\big(E_{\text{FP}} - E_0\big)}_{\text{phase err.} \,\ge\, 0}.
\label{eq:decomp}
\end{equation}
The measured gap $\Delta E = E_{\text{VMC}} - E_{\text{DMC}}$ is thus a rigorous, quantitative measure of the residual \emph{amplitude} error of the VMC state. A vanishing gap ($\Delta E \to 0$ within statistical resolution) demonstrates that the optimization has saturated the variational freedom permitted by its learned phase. The phase error, by contrast, is not probed by $\Delta E$: since the learned phase may itself be in error, a vanishing gap is necessary for the exact ground state but not sufficient.
 
In practice we perform importance-sampled FPDMC~\cite{Reynolds1982,Foulkes2001}: walkers undergo drift-diffusion moves guided by $\nabla \ln|\Psi_{\theta}|$, with branching weights determined by the local energy of $\hat H_{\text{eff}}$ in the standard short-time algorithm~\cite{Umrigar1993}. Because the projected state is the ground state of $\hat H_{\text{eff}}$, the mixed estimator $\langle \Psi_{\theta}|\hat H_{\text{eff}}|\Phi(\tau)\rangle / \langle \Psi_{\theta}|\Phi(\tau)\rangle$ converges to $E_{\text{FP}}$ without mixed-estimator bias~\cite{Foulkes2001}. Both the drift $\nabla\ln|\Psi_{\theta}|$ and the phase gradient $\nabla\Phi_{\theta}$ are obtained analytically from the network by automatic differentiation. We use an imaginary-time step $\Delta\tau = 0.001$ and a target population of 4092 walkers with standard population control; residual time-step and population-control biases are quantified in Appendix~\ref{app:timestep} and verified to lie below our statistical resolution~\cite{Umrigar1993,Zen2016}.

\section{\label{sec:results} Results}
The assessment framework of Sec.~\ref{sec:FPDMC} makes no reference to the internal structure of the trial state: it requires only a complex-valued trial wavefunction whose log-amplitude and phase gradients are available, and the Fermi Sets wavefunction supplies both by automatic differentiation. We apply the framework to two physically distinct systems: a finite quantum dot with explicitly broken time-reversal symmetry (Sec.~\ref{sec:qdot}), where multi-Landau-level exact diagonalization provides an independent reference, and the strongly correlated uniform electron gas (Sec.~\ref{sec:ueg}). The framework operates without modification in either case, and the diagnostic exhibits the same qualitative behavior across these deliberately contrasting problems---spatially confined versus extended, time-reversal-broken versus time-reversal-symmetric systems---showing that the observed behavior is not tied to a particular physical setting.

\subsection{\label{sec:qdot} Quantum dot in a magnetic field}
 
We first validate our diagnostic on a setup with an available reference: $N = 4$ electrons in a two-dimensional harmonic trap with GaAs parameters ($m^{*} = 0.067\,m_e$, $\hbar\omega_0 = 3.32$~meV, $g^{*} = -0.44$). The system is rotationally invariant, leading to a good quantum number $L$. Despite having no symmetry enforcement or built-in bias towards it, the network learns the rotational invariance by itself — Table~\ref{tab:lz_purity} illustrates how the converged states after energy minimization are almost exact $L$-eigenstates. Appendix~\ref{app:angmom} details the angular momentum decomposition of the neural network wavefunction.  

At $B=8$~T, the ground state is found to have $L=14$. For quantum dots, the filling factor is calculated as~\cite{manninen} 
\begin{equation}
    \nu =\frac{N(N-1)}{2L}.
\end{equation}
Thus, the $L=14$ ground state at $B=8$~T corresponds to a Jain-sequence composite fermion state at $\nu =3/7$. Additionally, we study this system at $B = 4$~T, where the ground state is found to have $L=6$, corresponding to the $\nu =1$ maximum density droplet (MDD). For both $B=8$~T and $4$~T, we benchmark against multi-Landau-level exact diagonalization, with the calculation detailed in Appendix~\ref{app:ed}. 

The significance of the quantum dot setup lies in the complicated phase structure: because the ground state is complex, the network has to learn a non-trivial winding phase. 
Table~\ref{tab:qdot} collects the VMC and DMC energies and residue gaps at both magnetic fields, and Table~\ref{tab:ed_conv} gathers the reference energies from exact diagonalization and DMC performed on a handcrafted composite-fermion ansatz~\cite{GucluNew}. 

\begin{figure}[t]
\includegraphics[width=\columnwidth]{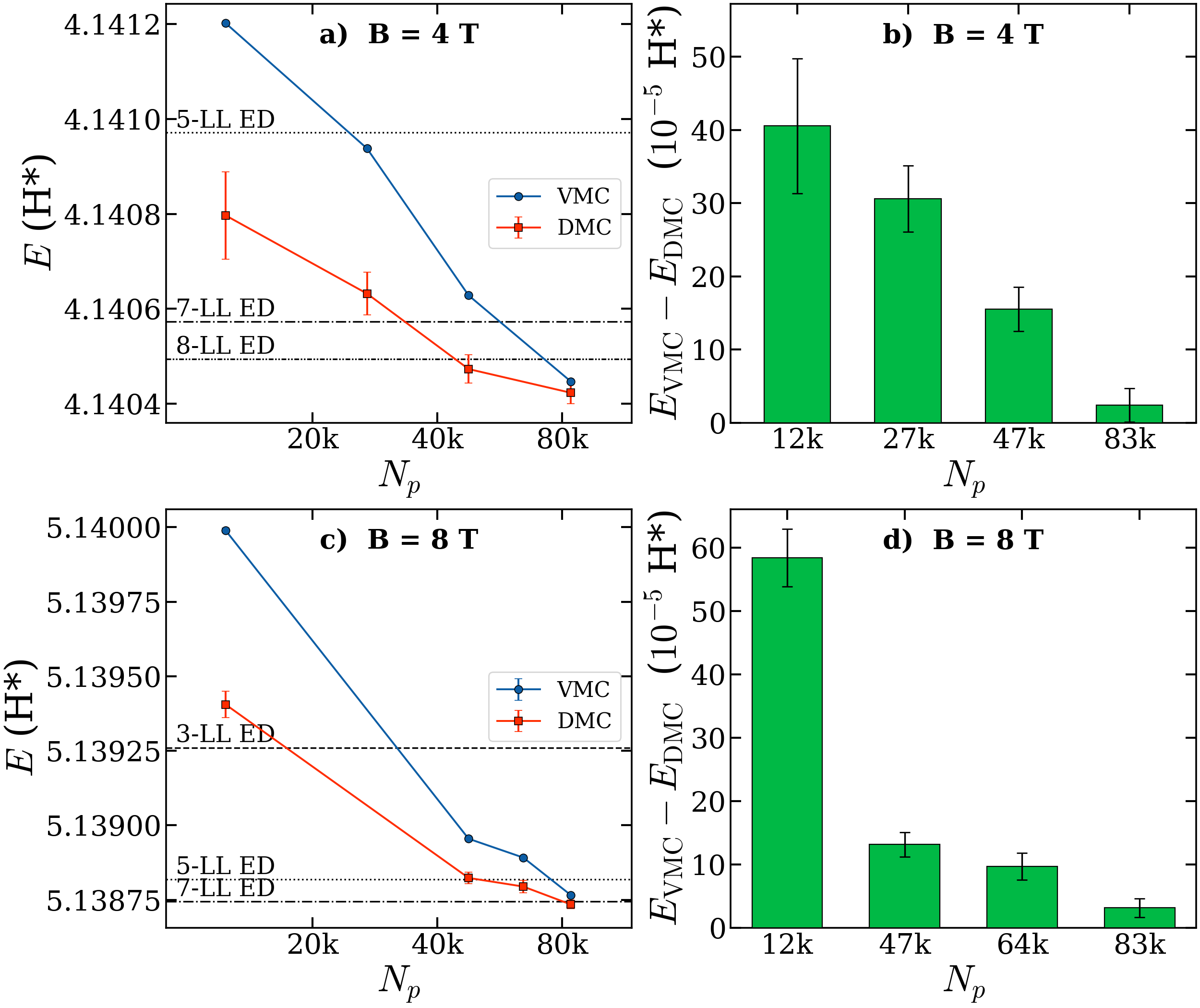}
\caption{\label{fig:qdot}%
Left: convergence of the VMC and DMC energies with the number of parameters for a) $B=8$~T and c) $B=4$~T. Right: Residue gap $\Delta E = E_{\text{VMC}} - E_{\text{DMC}}$. Error bars denote reblocked statistical uncertainties (Appendix~\ref{app:energy})}
\end{figure}

Figure~\ref{fig:qdot} shows that, as the network grows, the variational energy decreases monotonically toward the exact ground state value. For $B=4$~T, NN-VMC outperforms 8-LL ED, while for $B=8$~T, it reaches 7-LL ED, with DMC projection refining it to below the 7-LL benchmark.
That NN-VMC outperforms ED to extraordinary accuracy is verifiable evidence that the network is asymptotically approaching the exact ground state as the network size is increased. 
%With a combination of VMC and DMC, we reach energies even lower than 7LL ED for the $\nu=1$ setup at $B=4$~T. 

The projected DMC energy agrees with the variational VMC energy to within $0.02 \%$ for all network sizes. While DMC consistently improves VMC energy further, the additional gain from DMC or the residue gap shrinks from $561$ $\mu$H$^*$ to $25$ $\mu$H$^*$ for $B=8$~T, and from $405$ $\mu$H$^*$ to $24$ $\mu$H$^*$ for $B=4$~T. At $84$k parameters, we see that the residue gap has shrunk to the order of statistical uncertainty. That $\Delta E$ becomes statistically indistinguishable from zero at an increasing number of parameters is a clear indicator that %within the fixed phase error — which using ED has been ruled out to be very small - 
the network finds the %amplitude landscape of the 
true ground state with increasing precision. Thus, the exactness of a Fermi Sets state is only limited by the network size.  %eliminate all amplitude errors while discovering a highly complex phase manifold.

It is particularly noteworthy that Fermi Sets successfully finds the composite fermion state at $B=8$T, which cannot be described---even at a qualitative level---by any traditional Slater-Jastrow ansatz. Unlike Jastrow factors that only modify the amplitude, the Fermi Sets wavefunction uses fully general symmetric functions $\Omega_k(\mathbf r)$ that are complex-valued, which fundamentally alters the phase structure. This is crucial for its success in finding the composite fermion ground state.

\begin{table}[H]
\caption{\label{tab:qdot}%
Fixed-phase DMC assessment of Fermi Sets states for a quantum dot, in effective Hartrees (H$^*$), as a function of network parameter count $N_p$.
}
\begin{ruledtabular}
\begin{tabular}{lcccc}
\multicolumn{1}{c}{$B=8$~T} &
\multicolumn{1}{c}{$N_p\ (10^4)$} &
\multicolumn{1}{c}{$E_{\text{VMC}}$} &
\multicolumn{1}{c}{$E_{\text{DMC}}$} &
\multicolumn{1}{c}{$\Delta E$ ($10^{-4}\,$H$^*$)} \\
\colrule
{} & $1.2$ & 5.139966 & 5.139405(45) & 5.61 \\
{} & $4.8$ & 5.138962 & 5.138827(17) & 1.35 \\
{} & $6.4$ & 5.138895 & 5.138793(17) & 1.01 \\
{} & $8.4$ & 5.138760 & 5.138734(17) & 0.25 \\
\hline\hline
\multicolumn{1}{c}{$B=4$~T} &
\multicolumn{1}{c}{$N_p\ (10^4)$} &
\multicolumn{1}{c}{$E_{\text{VMC}}$} &
\multicolumn{1}{c}{$E_{\text{DMC}}$} &
\multicolumn{1}{c}{$\Delta E$ ($10^{-4}\,$H$^*$)} \\
\colrule
 {} & $1.2$ & 4.141202 & 4.140797(92) & 4.05 \\
 {} & $2.7$ & 4.140938 & 4.140632(45) & 3.06 \\
 {} & $4.8$ & 4.140628 & 4.140473(30) & 1.55 \\
 {} & $8.4$ & 4.140447 & 4.140423(23) & 0.24 \\
\end{tabular}
\end{ruledtabular}
\end{table}

\begin{table}[H]
\caption{\label{tab:ed_conv}%
Exact-diagonalization and composite-fermion DMC (CF+DMC)~\cite{GucluNew} energies of the quantum dot ground states as a function of the number of Landau levels retained. The bold row is our best Fermi Sets result (largest network, $N_p=8.4\times10^{4}$).
}
\begin{ruledtabular}
\begin{tabular}{lcc}
\multicolumn{1}{c}{$B=8$~T} &
\multicolumn{1}{c}{Method} &
\multicolumn{1}{c}{$E$ (H$^*$)} \\
\colrule
{} & 1-LL ED & 5.186000 \\
{} & 3-LL ED & 5.139260 \\
{} & 5-LL ED & 5.138819 \\
{} & 7-LL ED & 5.138745 \\
{} & CF + DMC~\cite{GucluNew} & 5.1394  \\
{} & \textbf{Fermi Sets} & \textbf{5.138760} \\
\hline \hline
\multicolumn{1}{c}{$B=4$~T} &
\multicolumn{1}{c}{Method} &
\multicolumn{1}{c}{$E$ (H$^*$)} \\
\colrule
{} & 1-LL ED & 4.295304 \\
{} & 3-LL ED & 4.143131 \\
{} & 5-LL ED & 4.140971 \\
{} & 7-LL ED & 4.140573  \\
{} & 8-LL ED & 4.140494 \\
{} & \textbf{Fermi Sets} & \textbf{4.140447} \\
\end{tabular}
\end{ruledtabular}
\end{table}

\subsection{\label{sec:ueg} Two-dimensional uniform electron gas}
We next study the ground state of the two-dimensional uniform electron gas (2D UEG)~\cite{GiulianiVignale2005}: interacting electrons embedded in a uniform, rigid, neutralizing positive background, characterized at zero temperature by the dimensionless density parameter $r_s = a/a_0$, where $a_0 = \hbar^2/me^2$ is the Bohr radius and $a = 1/\sqrt{\pi \rho}$ is the radius of the circle containing one electron on average at number density $\rho$. We simulate finite systems of $N$ electrons under periodic boundary conditions in a hexagonal simulation cell, with the long-range Coulomb interaction resummed by two-dimensional Ewald summation~\cite{Ewald1921,Tanatar1989,Fraser1996}; the cell geometry and further background on the model are collected in Appendix~\ref{app:ueg}. All electron-gas trial states are Fermi Sets wavefunctions with periodic input features (Sec.~\ref{sec:fermisets}), optimized as described in Appendix~\ref{app:vmc}.
 
We study the fully spin-polarized gas with $N = 16$ electrons at $r_s = 30$, which places the benchmark deep in the strongly correlated regime, immediately adjacent to the Wigner-crystallization density $r_s \simeq 31$ of the 2D UEG~\cite{Drummond2009}.

We fix the number of determinants at $N_d = 4$ and scale the hidden width of the network over three sizes, spanning roughly $N_p = 25\,414$ to $50\,339$ trainable parameters.

\begin{figure}[t]
\includegraphics[width=\columnwidth]{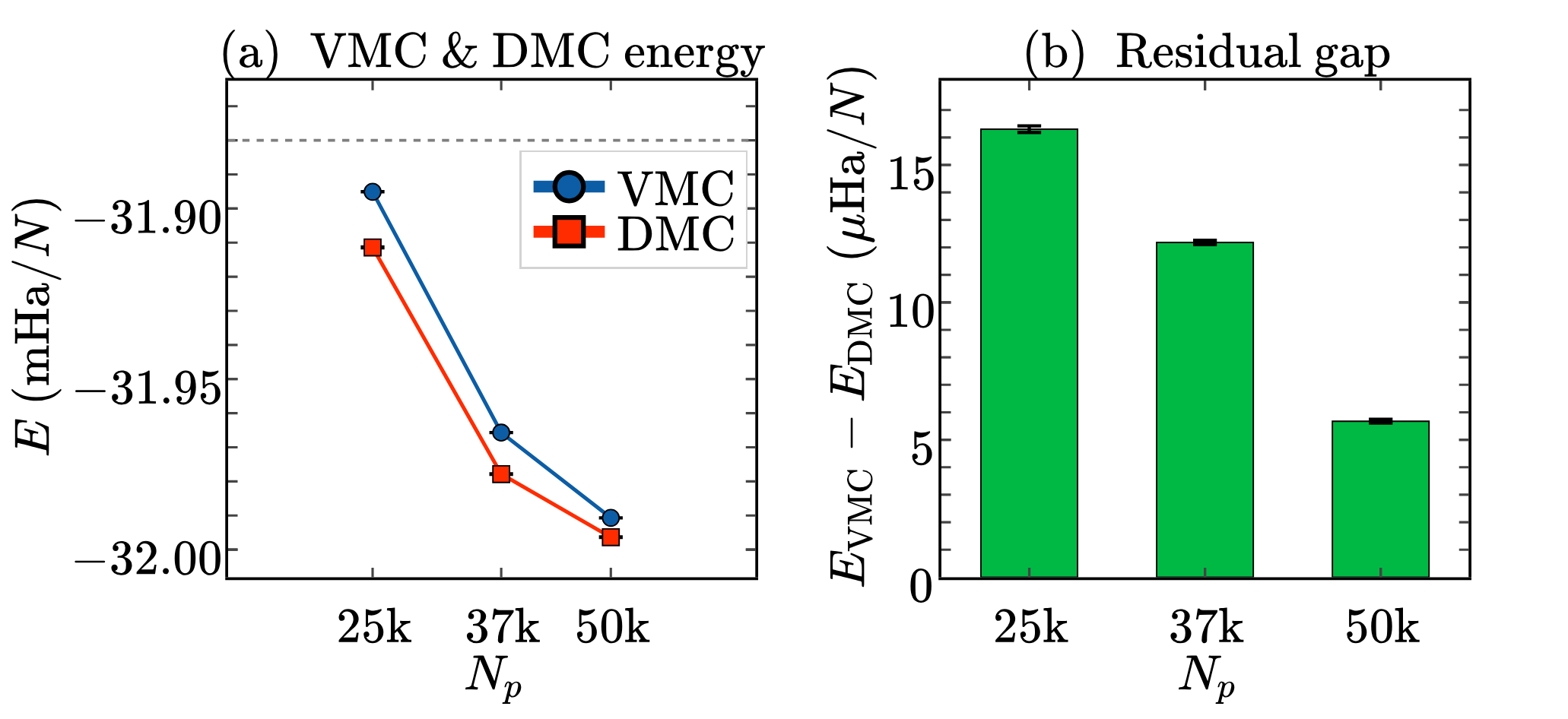}
\caption{\label{fig:dmc_2DEG}%
Convergence of the Fermi Sets states for the fully spin-polarized 2D UEG ($N=16$, $r_s=30$) with network parameter count $N_p$. (a)~VMC and fixed-phase DMC energies per electron. (b)~Residual gap $\Delta E = E_{\text{VMC}} - E_{\text{DMC}}$. Error bars denote reblocked statistical uncertainties  (Appendix ~\ref{app:energy}) and are smaller than the symbols where not visible.}
\end{figure}

Table~\ref{tab:ueg} and Fig.~\ref{fig:dmc_2DEG} collect the results of the size scan. Both the variational and projected energies descend monotonically as the network is scaled from 25k to 37k, and then to 50k parameters, with the DMC energy tracking the VMC energy closely at each step. 
For reference, we also list the energy of Tanatar and Ceperley (Ref.~\onlinecite{Tanatar1989}), obtained with a Slater--Jastrow trial wave function. Even our smallest network yields a variational energy below this benchmark, and the DMC energies improve upon it by up to $\sim 1.2\times10^{-4}$~Ha per electron at the largest network size.
Because the fixed-phase energy depends on the phase of the trial state alone (Sec.~\ref{sec:FPDMC}), the steady descent of $E_{\text{DMC}}$ is direct evidence that the learned phase improves systematically with capacity.

\begin{table}[H]
\caption{\label{tab:ueg}%
Fixed-phase DMC assessment of Fermi Sets states for the fully
spin-polarized 2D UEG ($N = 16$, $r_s = 30$), as a function of network
parameter count $N_p$. Energies are per electron, in Hartree;
statistical uncertainties in the last digits are given in parentheses
(Appendix~\ref{app:energy}).}
\begin{ruledtabular}
\begin{tabular}{cccc}
\multicolumn{1}{c}{$N_p\ (10^4)$} &
\multicolumn{1}{c}{$E_{\text{VMC}}$ (Ha/$N$)} &
\multicolumn{1}{c}{$E_{\text{DMC}}$ (Ha/$N$)} &
\multicolumn{1}{c}{$\Delta E$ ($10^{-5}\,$Ha/$N$)} \\
\colrule
$2.5$ & $-0.031895091(12)$ & $-0.03191139(12)$ & $1.630(12)$ \\
$3.7$ & $-0.031965678(6)$ & $-0.03197786(8)$ & $1.218(8)$ \\
$5.0$ & $-0.031990687(3)$ & $-0.03199636(7)$ & $0.567(7)$ \\
\colrule
\multicolumn{1}{l}{Ref.~\cite{Tanatar1989}} &
$-0.031880(15)$ & \multicolumn{1}{c}{---} & \multicolumn{1}{c}{---} \\
\end{tabular}
\end{ruledtabular}
\end{table}

The residue gap decreases monotonically in parallel, from $\Delta E = 1.630(12)\times10^{-5}$ to $0.567(7)\times10^{-5}$~Ha per electron, though it remains statistically resolved at the largest network. The residual amplitude error is thus systematically suppressed as capacity grows, consistent with the proven improvability of the ansatz~\cite{Fu2026FermiSets}, and our lowest energy, $E_{\text{DMC}} = -0.03199636(7)$~Ha per electron, is a variational upper bound on the ground-state energy at this system size.

\section{\label{sec:discussion} Discussion and conclusions}
 
We have introduced a general framework that integrates complex-valued neural-network wavefunctions with fixed-phase diffusion Monte Carlo, using the projection both to improve optimized variational energy and to certify their accuracy: in both systems studied, the residual gap $\Delta E = E_{\text{VMC}} - E_{\text{DMC}} \ge 0$ shrinks systematically with network
capacity.

Two extensions follow naturally. Combined with twist averaging and finite-size extrapolation, the $\Delta E$ diagnostic would carry certified accuracy to the thermodynamic limit; for the 2D UEG, this would allow us to locate the Wigner-crystallization boundary more accurately.  %at $r_s \simeq 31$~\cite{Drummond2009}, immediately adjacent to the density studied here. 
Because the fixed-phase constraint accommodates arbitrary complex trial states~\cite{Ortiz1993,Bolton1996}, the framework also applies without modification to fractional quantum Hall liquids~\cite{Teng2025FQH, abouelkomsan2026, gattu2025, fadon2025} and fractional Chern insulators in moir\'e materials~\cite{li25, luo25}.

\begin{acknowledgments}
%We thank Ahmed Abouelkomsan for helpful discussions and related collaboration. 
This work was supported by Air Force Office of Scientific Research under award number FA2386-24-1-4043. SP acknowledges support from the MIT Undergraduate Research Opportunities Program (UROP). YL, HL and LF are grateful for the support from MISTI Global Seed Funds. T.-R.C. was supported by National Science and Technology Council (NSTC) in Taiwan (Program No. NSTC 113-2124-M 006-009-MY3). H.L. acknowledges the support by the National Science and Technology
Council (NSTC) in Taiwan under grant number NSTC 114-2112-M-001-055-MY3. LF was supported in part by a Simons Investigator Award from the Simons Foundation. We thank the MIT Engaging and the NCHC Nano4 cluster for providing the computational resources for the NN-VMC and DMC simulations. 

\end{acknowledgments}

\appendix 
\section{\label{app:vmc} Variational Monte Carlo details}
According to the Rayleigh--Ritz variational principle, the expectation value of the Hamiltonian $\hat{H}$ evaluated with the trial wavefunction $\Psi_{\theta}$ provides a strict upper bound to the exact ground-state energy, $E_0$:
\begin{equation}
E_{\text{VMC}}(\theta) = \frac{\langle \Psi_{\theta} | \hat{H} | \Psi_{\theta} \rangle}{\langle \Psi_{\theta} | \Psi_{\theta} \rangle} \ge E_0.
\end{equation}
In the VMC framework, this multidimensional integral is recast as a statistical expectation value over the probability distribution $p(\mathbf{R}) \propto |\Psi_{\theta}(\mathbf{R})|^2$. The energy can thus be evaluated as a Monte Carlo average of the local energy,
\begin{equation}
E_{\text{VMC}}(\theta) = \mathbb{E}_{\mathbf{R} \sim p(\mathbf{R})} \left[ E_L(\mathbf{R}; \theta) \right],
\qquad
E_L = \frac{\hat{H}\Psi_{\theta}}{\Psi_{\theta}}.
\end{equation}
For a complex trial state the local energy is itself complex-valued; because $\hat H$ is Hermitian, its imaginary part averages to zero and only the real part contributes to $E_{\text{VMC}}$.
 
The electron configurations $\mathbf{R}$ are sampled with the Metropolis--Hastings algorithm~\cite{Metropolis1953,Hastings1970}: at each optimization step, an ensemble of walkers explores configuration space guided by the probability density $|\Psi_{\theta}(\mathbf{R})|^2$.

To drive the trial wavefunction toward the ground state, the parameters are updated along the energy gradient, which for a complex wavefunction with real parameters takes the standard covariance form~\cite{BeccaSorella2017,Pfau2020FermiNet} 
\begin{equation}
\nabla_{\theta} E_{\text{VMC}} = 2\,\mathrm{Re}\,
\mathbb{E}_{\mathbf{R}\sim p}\!\left[
\big(E_L-E_{\text{VMC}}\big)\,\nabla_{\theta}\log\Psi^{*}_{\theta}
\right],
\label{eq:grad}
\end{equation}
evaluated over the same Monte Carlo samples.
 
 For neural-network ansatzes, plain gradient descent is inefficient owing to the ill-conditioned geometry of the parameter space; the gradient is therefore preconditioned with the quantum geometric (Fisher) metric, as in stochastic reconfiguration~\cite{Sorella1998} and its scalable Kronecker-factored approximation K-FAC~\cite{MartensGrosse2015}, the latter adapted to neural wavefunctions in Ref.~\cite{Pfau2020FermiNet}. In this work we optimize with K-FAC  for 500k iterations using a batch of 4{,}096 walkers and initial learning rate start from $10^{-3}$.

\section{\label{app:backbone}Network architecture for the quantum dot problem}

The single-particle orbitals $\phi_i^{k}$ and the symmetric factors $\Omega_k$ of the Fermi
Sets wavefunction, Eq.~(\ref{eq:fermisets}), are implemented for the quantum dot problem as
follows. Throughout we use $N_d=2$ determinants and a $\tanh$ nonlinearity. The backbone
width $d$ is the single knob we sweep, from $d=24$ to $d=64$, to vary the network size.

The symmetric factors are built from features produced by a backbone
$h:\mathbb{R}^{3N}\to\mathbb{R}^{N\times d}$. Each electron is first embedded from its
single-particle inputs $\bm g_i=(x_i,y_i,r_i^2)$,
\begin{equation}
h^{(0)}_i=W_{\mathrm{in}}\bm g_i,\qquad W_{\mathrm{in}}\in\mathbb{R}^{d\times3}.
\end{equation}
The radial coordinate $r_i^2$ is supplied alongside $(x_i,y_i)$ because it is the term that
enters the confinement, so providing it directly spares the network from synthesizing it
from the Cartesian inputs. Each block $b=1,...,L_b=3$ applies multi-head self-attention
followed by a per-electron MLP,
\begin{align}
\tilde h^{(b)}_i &= h^{(b-1)}_i+\mathrm{MHA}\big(\tanh h^{(b-1)}\big)_i,
\label{eq:mha}\\
h^{(b)}_i &= \tilde h^{(b)}_i+W_2\tanh\big(W_1\tilde h^{(b)}_i\big),
\qquad W_1,W_2\in\mathbb{R}^{d\times d},
\label{eq:mlp}
\end{align}
where the multi-head attention with $n_h$ heads reads
\begin{equation}
\begin{split}
\mathrm{MHA}(x)_i &= W_O\bigoplus_{a=1}^{n_h}\sum_{j} A^{a}_{ij}\,W_V^a x_j,\\
A^{a}_{ij} &= \mathrm{softmax}_j\left(
\frac{(W_Q^a x_i)\cdot(W_K^a x_j)}{\sqrt{d/n_h}}\right).
\end{split}
\label{eq:attn}
\end{equation}
with $\oplus$ denoting concatenation. We use $n_h=8$ attention heads, except for $d=36$,
where $n_h=6$ is used instead. The self-attention couples every electron to all others, and
is the origin of the backflow correlations, which in our construction are carried entirely
by the symmetric factors.

To construct the symmetric factor, the pooled collective coordinate
\begin{equation}
\bm{\xi}=\sum_{i=1}^{N}h^{(L_b)}_i\in\mathbb{R}^{d}
\label{eq:pooling}
\end{equation}
is passed to a complex MLP $\Omega_k:\mathbb{R}^{d}\to\mathbb{C}$,
\begin{equation}
\Omega_k(\bm\xi) = \big(W_2^{\mathrm{re}} + i W_2^{\mathrm{im}}\big)\tanh(W_1\bm\xi)
\end{equation}
with $W_1\in\mathbb{R}^{d\times d}$ and
$W_2=(W_2^{\mathrm{re}}, W_2^{\mathrm{im}})^T\in\mathbb{R}^{2\times d}$.

The orbitals are constructed from $\bm g_i$ using a single-layer MLP
$\bm\eta_i=\tanh(W_{\phi}\bm g_i)$ of width $m=8$. The $\bm\eta_i$ are turned into complex
orbitals using a learned projection
\begin{equation}
\phi_i^{k}(\mathbf x_j)=\bm w^{k,\mathrm{re}}_{i}\cdot\bm\eta_j
+i\,\bm w^{k,\mathrm{im}}_{i}\cdot\bm\eta_j,
\qquad \bm w^{k,\mathrm{re}}_i,\bm w^{k,\mathrm{im}}_i\in\mathbb{R}^{m}.
\label{eq:orbital}
\end{equation}
With a single layer of width $m=8$, the orbitals provide a simple node structure, and the
entire burden of representing the many-body correlation falls on the symmetric factors
$\Omega_k$.

\section{\label{app:params}Parameter count for the quantum dot problem}
Almost all parameters live in the attention backbone. Collecting every weight and bias by
its scaling in the width $d$, the total parameter count is exactly
\begin{equation}
\begin{split}
N_p &= (c_bL_b+N_d)\,d^2 + (c_bL_b+3N_d+4)\,d + C \\
    &\phantom{{}={}} = 20\,d^2 + 28\,d + 166,
\end{split}
\label{eq:nparams}
\end{equation}
with $c_b=6$, $L_b=3$, $N_d=2$.

The $d^2$ term counts the weight matrices that scale with width: the $c_b=6$ matrices of
each of the $L_b=3$ backbone blocks (the four attention projections $W_Q,W_K,W_V,W_O$ and
the two MLP layers $W_1,W_2$), plus the $d\times d$ hidden layer of each of the $N_d$
symmetric-factor MLPs $\Omega_k$. The linear term collects the input embedding
$W_{\mathrm{in}}\in\mathbb{R}^{d\times3}$, the $2\times d$ output projections of the $\Omega_k$,
and all per-layer biases. The constant $C=166$ gathers everything independent of $d$: the
single-body orbital branch (read-outs $2N_dNm=128$ and embedding $4m=32$), the
symmetric-factor output biases ($2N_d=4$), and the two envelope and Jastrow scalars.

To leading order $N_p\sim(c_bL_b+N_d)\,d^2=20\,d^2$, quadratic in the width $d$ and linear in
the depth $L_b$, with the backbone and symmetric factors together holding about $90\%$ of
the total. We vary the network size by sweeping $d$ over five values $[24;\ 36;\ 48;\ 56;\ 64]$, spanning
$N_p\in[12,358;\ 27,094; \ 47,590; \ 64,454; \ 83,878]$.

\section{\label{app:ed}Exact diagonalization}

To provide an independent benchmark we perform our own exact diagonalization in a truncated single-particle basis. The natural
basis is the set of Fock--Darwin states $|n,m\rangle$ of the two-dimensional harmonic
oscillator in a magnetic field, labelled by the Landau-level index $n$ and angular momentum
$\ell=m-n$, with single-particle energies
$\varepsilon(n,m)=(n+\tfrac12)\omega_+ + (m+\tfrac12)\omega_-$ where
$\omega_\pm=\omega\pm\omega_c/2$ and $\omega=\sqrt{\omega_0^2+\omega_c^2/4}$. We build
the many-body basis from all Slater determinants of $N$ such orbitals within a fixed
total angular-momentum sector, and diagonalize the full Hamiltonian in that sector. The
two-body Coulomb matrix elements are evaluated in momentum space using the closed-form
expressions for the Fock--Darwin form factors
$\langle a|e^{i\bm q\cdot\bm r}|c\rangle$, and the lowest eigenvalue is obtained by sparse
(Lanczos) diagonalization.

\section{\label{app:ueg} The 2D uniform electron gas: background and simulation cell}
 
The 2D UEG is a paradigmatic many-body model whose accurate characterization underpins the construction of exchange-correlation functionals in density functional theory~\cite{Tanatar1989,Attaccalite2002}. Quantum Monte Carlo studies place the transition from the paramagnetic fluid to a triangular Wigner crystal at $r_s = 31(1)$~\cite{Tanatar1989,Kwon1993,Attaccalite2002,Drummond2009}. Neural-network wavefunctions have been applied to the electron gas in both three and two dimensions~\cite{Cassella2023HEG,Wilson2023HEG,Smith2024UEG}, making it a natural and demanding testbed for the present framework.
 
The hexagonal simulation cell---the primitive cell of a triangular lattice, whose Wigner--Seitz cell is a regular hexagon---is defined by the primitive lattice vectors $\mathbf{L}_1 = L(1, 0)$ and $\mathbf{L}_2 = L(-1/2, \sqrt{3}/2)$. For the isotropic UEG this geometry is preferable to a square cell: its higher rotational symmetry mitigates anisotropic finite-size effects, and it accommodates the triangular Wigner crystal without distortion. Equating the cell area, $A = (\sqrt{3}/2)L^2$, to the total area occupied by the $N$ electrons, $A = N\pi (r_s a_0)^2$, fixes the box size:
\begin{equation}
L = r_s a_0 \sqrt{\frac{2\pi N}{\sqrt{3}}}.
\label{eq:boxsize}
\end{equation}
Interparticle distances are evaluated in the minimum-image convention~\cite{AllenTildesley1987}. 

The Hamiltonian of the interacting $N$-electron system reads
\begin{equation}
\hat{H} = -\frac{1}{2r_s^2} \sum_{i=1}^N \nabla_i^2 + \frac{1}{r_s} \sum_{1 \le i < j \le N} \frac{1}{|\mathbf{r}_i - \mathbf{r}_j|} + \text{const},
\label{eq:hamiltonian}
\end{equation}
where $r_s$ enters in the prefactor as detailed by Ref.~\cite{Ortiz1993,Bolton1996}. The first term in Eq.~\ref{eq:hamiltonian} is the kinetic energy, the second the Ewald-resummed electron-electron repulsion, and the constant is the Madelung energy of the periodic cell: the uniform background cancels the divergent zero-momentum ($\mathbf{G}=0$) contribution exactly~\cite{Fraser1996}. 

\section{\label{app:energy} Energy evaluation and standard error of the mean}
 
For both the VMC and FPDMC estimates we perform inference runs of $500,000$ Monte Carlo steps with an ensemble of 4{,}096 parallel walkers. Statistics are accumulated over the final $M=2^{18}$ steps of each trajectory, with the preceding segment discarded as equilibration; for FPDMC this burn-in removes any systematic bias from the initial imaginary-time transient. One ensemble-averaged local-energy measurement is recorded per step, yielding $M = 2^{18}$ measurements per run.

All statistical uncertainties are quoted as the standard error of the mean (SEM). For statistically independent samples the SEM would be $\sigma(E_L)/\sqrt{M}$, with $\sigma(E_L)$ the standard deviation of the recorded measurements; consecutive Monte Carlo configurations are, however, autocorrelated, so this naive estimate underestimates the true uncertainty. We therefore obtain the SEM by reblocking analysis~\cite{Flyvbjerg1989}: the local-energy measurements are averaged over blocks, and the SEM is then calculated over the block-averaged energy values. The block size is systematically increased until the estimated SEM converges to a plateau, and the plateau SEM is quoted as the uncertainty on both $E_{\text{VMC}}$ and $E_{\text{DMC}}$. Since the VMC and FPDMC production runs are statistically independent, the SEM of the residue gap $\Delta E = E_{\text{VMC}} - E_{\text{DMC}}$ is obtained by combining the two SEMs in quadrature.

\section{\label{app:angmom}Angular-momentum decomposition}

The quantum dot setup is rotationally invariant, $[\hat H,\hat L_z]=0$, so the angular momentum $L_z$ is a good quantum number. We extract each state's angular momentum content by rotation
projection. Under a rotation $\mathcal R_\phi$ by angle $\phi$, an
$L_z$-eigenstate obeys $\psi(\mathcal R_\phi R)=e^{iL\phi}\psi(R)$, so the overlap
$\langle \Psi_{\theta} | \mathcal R_\phi | \Psi_{\theta} \rangle = \sum_L |c_L|^2 e^{iL\phi}$ is a Fourier series in $\phi$ whose
coefficients are precisely the sector weights $|c_L|^2$. We decompose the state by a  discrete Fourier transform over $N_\phi$ rotation angles. 

At $B=8$~T the ground state lies in the $L_z=14$ sector, while the $B=4$~T ground state lies in the $L_z=6$ sector, as determined by ED, and confirmed by VMC in Table\ref{tab:lz_purity}. We note that NN-VMC independently learns the rotational invariance, with all network sizes converging to less than $0.1\%$ weight outside their expected $L_z$-sector.

\begin{table}[b]
\caption{\label{tab:lz_purity}%
Angular-momentum purity of the Fermi Sets states as a function of
network parameter count $N_p$.
}
\begin{ruledtabular}
\begin{tabular}{lcc}
\multicolumn{1}{c}{$B=8$~T} &
\multicolumn{1}{c}{$N_p$} &
\multicolumn{1}{c}{$|c_{L=14}|^{2}$ (\%)} \\
\colrule
{} & $1.2\times10^{4}$ & 99.9160 \\
{} & $4.8\times10^{4}$ & 99.9944 \\
{} & $6.4\times10^{4}$ & 99.9955 \\
{} & $8.4\times10^{4}$ & 99.9990 \\
\hline\hline
\multicolumn{1}{c}{$B=4$~T} &
\multicolumn{1}{c}{$N_p$} &
\multicolumn{1}{c}{$|c_{L=6}|^{2}$ (\%)} \\
\colrule
{} & $1.2\times10^{4}$ & 99.9631 \\
{} & $2.7\times10^{4}$ & 99.9776 \\
{} & $4.8\times10^{4}$ & 99.9903 \\
{} & $8.4\times10^{4}$ & 99.9956 \\
\end{tabular}
\end{ruledtabular}
\end{table}

\section{\label{app:timestep} Time-step and population-control bias}
The short-time factorization of the FPDMC propagator introduces a systematic bias that vanishes as $\Delta\tau \to 0$~\cite{Umrigar1993}. To keep this bias small at practical time steps we employ the size-consistent drift-limiting and branching modifications of Zen \emph{et al.}~\cite{Zen2016}, including the associated local-energy cutoff (parameter $\alpha$).

\begin{figure}[t]
\includegraphics[width=\columnwidth]{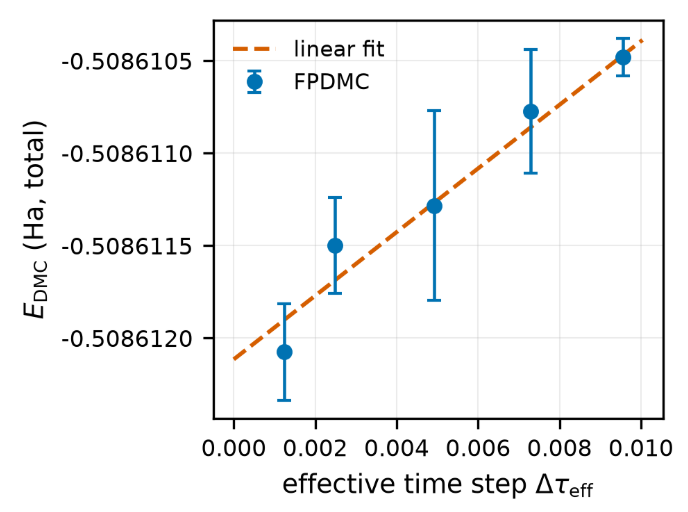}
\caption{\label{fig:timestep}%
Time-step dependence of the FPDMC energy for the fully spin-polarized
2D UEG ($N = 16$, $r_s = 30$; production network). The line is a fit
used to extrapolate to $\Delta\tau \to 0$; the production time step is
$\Delta\tau_0 = 0.001$. Error bars denote the reblocked standard
error of the mean (Appendix~\ref{app:energy}).}
\end{figure}

To verify that the production time step is small enough, we repeat the FPDMC projection for the 2D UEG ($N = 16$, $r_s = 30$; $N_d = 4$, production network) at a series of time steps bracketing the production value $\Delta\tau_0 = 0.001$. The total projected imaginary time is held fixed by scaling the number of steps inversely with $\Delta\tau$, so that the statistical uncertainties (Appendix~\ref{app:energy}) are comparable across the scan. Figure~\ref{fig:timestep} shows $E_{\text{DMC}}(\Delta\tau)$ together with a fit used to extrapolate to $\Delta\tau \to 0$; the extrapolated energy differs from the production result at $\Delta\tau_0$ by less than the statistical uncertainty, confirming that the time-step bias at $\Delta\tau_0 = 0.001$ is negligible on the scale of our error bars.

\newpage
\bibliography{apssamp}
 
\end{document}